\documentclass[fleqn,usenatbib]{mnras}

\usepackage{newtxtext,newtxmath}

\usepackage[T1]{fontenc}

\DeclareRobustCommand{\VAN}[3]{#2}
\let\VANthebibliography\thebibliography
\def\thebibliography{\DeclareRobustCommand{\VAN}[3]{##3}\VANthebibliography}

\usepackage{graphicx}	
\usepackage{amsmath}	
\usepackage{color}	    

\newcommand* {\kms}         {\mbox{km\,s$^{-1}$}}
\newcommand* {\kmskpc}      {\mbox{km\,s$^{-1}$\,kpc$^{-1}$}}
\newcommand* {\Msun}        {\mbox{M$_\odot$}}

\newcommand* {\diff}        {\mathrm{d}}
\renewcommand*{\vec}[1]     {\boldsymbol{#1}}
\newcommand* {\hvec}[1]     {\hat{\vec{#1}}}
\newcommand* {\tens}[1]     {\vec{\mathsf{#1}}}

\newcommand* {\sub}[2]      {{#1}_{\mathrm{#2}}}

\title[The Milky-Way Cepheid warp]
{A twisted and precessing Cepheid warp in the outer Milky Way disc}

\author[Dehnen, Semczuk \& Sch{\"o}nrich]{%
Walter Dehnen\href{http://orcid.org/0000-0001-8669-2316}{\includegraphics[width=9pt]{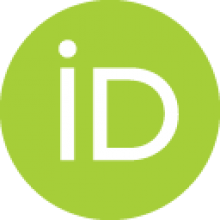}},$^{\!\!1,2}$
Marcin Semczuk\href{http://orcid.org/0000-0002-8191-8918}{\includegraphics[width=9pt]{orcid-ID}},$^{\!\!2}$ and
Ralph Sch{\"o}nrich\href{http://orcid.org/0000-0002-4236-3091}{\includegraphics[width=9pt]{orcid-ID}},$^{\!\!3}$
\smallskip
\\
$^1$ Astronomisches Recheninstitut, Zentrum f{\"u}r Astronomie der Universit{\"a}t Heidelberg, M{\"o}nchhofstra\ss{}e 12-14, 69120, Heidelberg, Germany \\
$^2$ School for Physics and Astronomy, University of Leicester, University Road, LE1 7RH, Leicester, UK \\
$^3$ Mullard Space Science Laboratory, University College London, Holmbury St.~Mary, Dorking, Surrey, RH5 6NT, UK
}

\date{Accepted XXX. Received YYY; in original form ZZZ}

\pubyear{2023}

\begin{document}

\defcitealias{Gaia22}{G22}
\defcitealias{D22}{D22}

\label{firstpage}
\pagerange{\pageref{firstpage}--\pageref{lastpage}}
\maketitle

\begin{abstract}
We examine the Galactic warp in a sample of all classical Cepheids with Gaia DR3 radial velocity. In each radial bin, we determine (1) the inclined plane normal to the mean orbital angular momentum of the stars and (2) that best fitting their positions. We find no warping inside $R\approx11\,$kpc; for larger $R$ the disc is increasingly inclined, reaching $i\sim3\degr$ at $R\geq 14\,$kpc. With larger $R$ the azimuth of the warp's ascending node shifts from $\sub{\varphi}{lon}\approx-15\degr$ at 11\,kpc by about 14\degr/kpc in the direction of Galactic rotation, implying a leading spiral of nodes, the general behaviour of warped galaxies. From the method of fitting planes to the positions we also obtain $\sub{\dot{\varphi}}{lon}$ and find prograde precession of $\sub{\dot{\varphi}}{lon}\sim12\,\kmskpc$ at 12\,kpc decreasing to $\sim6\,\kmskpc$ at 14\,kpc and beyond. This would unwind the leading spiral of nodes in $\sim100\,$Myr, suggesting that our instantaneous measurements of $\sub{\dot{\varphi}}{lon}$ reflect  transient behaviour. This is consistent with existing simulations, which show oscillations in $\sub{\dot{\varphi}}{lon}$ overlaying a long-term retrograde differential precession which generates the leading spiral of nodes.
\end{abstract}




\section{Introduction}
At least half of all edge-on spiral galaxies appear to be warped \citep*{SanchezSaavedraEtAl1990, Bosma1991, GarciaRuiz2002}. Since a warp is not detectable from all edge-on orientations, this implies that in fact most disc galaxies are warped. The simplest model conceptualizes a warp as a sequence of nested rings in circular motion, each tilted with respect to the flat inner galaxy. \cite{Briggs1990} fitted such tilted-ring models to the \ion{H}{I} distribution of spiral galaxies and found the warp typically begins near the Holmberg radius and reaches tilt angles of 5-20\degr, though values up to 90\degr\ (for polar-ring galaxies) also occur. As soon as an appreciable tilt is reached, its orientation can be measured, and \cite{Briggs1990} found them to vary with radius such that the line of nodes forms a leading spiral (`Briggs' rule'). 

The \ion{H}{I} warp of the Milky Way is clearly visible in 21cm emission \citep*{Burke57, Westerhout1957, Henderson1982}. Using kinematic distances, \cite{Burton1988} found it to commence at $R=10\,$kpc with increasing tilt (3\degr\ at 16\,kpc, which for the Milky Way corresponds the Holmberg radius), but also to being lopsided outside 14.5\,kpc (from \citealt{Burton1988} Fig.~7.19 after correcting for his too large value for $R_0$). Whether or not the Galactic \ion{H}{I} warp is twisted with a leading spiral of nodes, as for external galaxies, is difficult to assess, since (i) the lopsidedness complicates the picture and (ii) the line of nodes is close to the anti-centre direction, where kinematic distances are unavailable. \cite{Burton1988}'s analysis suggests a twist (in the same sense as seen in external galaxies) inside 14.5\,kpc, while \citeauthor{Levine06} (\citeyear{Levine06}, also using kinematic distances) concluded that the data are consistent with the line of nodes being in the anti-centre direction at all radii.

The Milky-Way warp is also manifest in the distribution of stars and dust across the sky \citep[][among others]{Djorgovski89, Freudenreich1994, Drimmel01, LC02, Reyle2009}, but for a quantitative analysis the distances to the tracers are required. Accurate astrometric parallax measurements only reach as far as 100\,pc (Hipparcos) and 3-5\,kpc (Gaia, depending on magnitude), i.e.\ the (extended) Solar neighbourhood. Nonetheless, the stellar warp has been identified in this region \citep{Chrobakova22}, in particular through the variation of the mean vertical motion with position, from Hipparcos \citep{Dehnen98, Drimmel2000} and more clearly from Gaia data \citep{SD18, Poggio18, Poggio20, Cheng20}. For more extended studies of the stellar warp, less accurate photometric distances can be utilized \citep{Lopez-Corredoira2014, RG19, Cheng20, Li2020}, but the large systematic distance uncertainties (e.g. from misclassification and extinction) risk bungling properties inferred for the warp. More recently, sufficiently large samples of Galactic Cepheids with accurate distances became available. These provide a wide view over the young stellar disc on our side of the Milky Way and clearly show the warp structure \citep{Skowron2019A, Skowron2019B, Chen2019, Lemasle2022}.

Most of these studies represent the mean vertical displacement of stars via an azimuthal Fourier series,
\begin{align}
    \label{eq:Z:Fourier}
    \bar{z}(R,\varphi) = Z_0 + Z_1 \sin(\varphi-\psi_1) + Z_2 \sin2(\varphi-\psi_2)
\end{align}
with amplitudes $Z_m$ and phases $\psi_m$ that are functions of radius $R$. An inclined plane with inclination $i\ll90\degr$ produces an $m=1$ component with amplitude $Z_1=R\tan i$ and line of nodes $\sub{\varphi}{lon}=\psi_1$; $m=0$ corresponds to an overall vertical offset of the disc and $m=2$ a saddle-shaped bend, rendering the warp lopsided. Many of the aforementioned studies restrict themselves to the $m=1$ component, limit the $\psi_m$ to constants, and the $Z_m(R)$ to simple functional forms. Their findings for the stellar warp agree reasonably well with those found for \ion{H}{I}: the warp begins at $\sim11\,$kpc and reaches tilts of $\sim3\degr$ at 16\,kpc with a line of nodes close to the anti-centre direction, apparently without appreciable twist.

A somewhat more detailed analysis was performed by \cite{Chen2019}, who in addition to a general fit of an $m=1$ model, split their sample of Cepheids into radial bins and fitted a titlted ring ($m=1$) to each. They found the line of nodes to vary radially with $\sub{\varphi}{lon}\sim0\degr$ at 12.5\,kpc and increasing to $\sim20\degr$ at $R>14.5\,$kpc\footnote{\cite{Chen2019} also found $\sub{\varphi}{lon}$ to increase inwards from $12.5\,$kpc (forming a trailing spiral), where the tilt is generally small. However, this result is less reliable, because at such small inclinations the best-fitting tilted ring is more strongly affected by dust obscuration. While not reflected in their error bars, this makes the line of nodes highly uncertain.}, thus forming a leading spiral and adhering to Briggs' rule.

For a warp consisting of tilted rings in near-circular motion, hereafter termed `simple' warp, (i) only the $m=1$ component of the series~\eqref{eq:Z:Fourier} has significant amplitude and (ii) the warp undergoes nodal precession with rate $\dot{\psi}_1=\sub{\Omega}p$, where $\sub{\Omega}p\equiv\Omega_\varphi-\Omega_z$ is the precession rate of individual stellar orbits ($\Omega_\varphi$ and $\Omega_z$ are the orbital frequencies for azimuthal rotation and vertical oscillation). For oblate systems $\Omega_\varphi \lesssim \Omega_z$, such that a simple warp precesses slowly backwards (retrograde) and, since $|\sub{\Omega}p|$ decreases outwards, differential precession naturally winds up the line of nodes into a leading spiral.

However, there are clear indications that the Milky-Way warp is not simple. First, the \ion{H}{I} distribution shows strong lopsidedness, implying a significant $m=2$ component. Second, when interpreting the vertical stellar motions in terms of a simple warp model, \cite{Poggio20} and \cite{Cheng20} found precession rates of $10.9$ and 13.6\,\kmskpc from samples of giants and the general stellar population, respectively, indicating fast prograde precession ($\Omega_\varphi\sim20\,\kmskpc$ at the relevant radii), instead of slow backward precession (note, though, that using similar analysis \citealt{Wang2020} and \citealt{Chrobkova2021} found their data to be consistent with a static non-precessing warp). All these studies combined highly uncertain astrometric parallaxes with photometric distances, that are known to be prone to systematic biases which dominate over the statistical uncertainties and are hard to quantify (\citealt*{Schoenrich2011}; \citealt{Williams2013}). When using Cepheids, these problems are much reduced and the errors largely limited to statistical uncertainties, which are larger due to smaller sample size, but fully quantifiable.

In the present study of the Cepheid warp, we are improving on previous studies in several ways. First, we consider the distribution over the angular-momentum directions of the stars, i.e.\ their individual instantaneous orbital planes. This is similar to the pole-count map employed by \cite{RG19} for OB and RGB star samples, except that we do not need to marginalise over the radial velocity, since these are known for all Cepheids in our sample. Second, we fit precessing tilted planes to the distributions of the Cepheids. The orientation of the respective planes is conventionally described in terms of their inclination $i$ and the Galactocentric azimuth $\sub{\varphi}{lon}$ of the line of nodes. In case of the tilted planes fitted to the Cepheid positions, we also obtain the time derivatives of these, i.e.\ $\diff i/\diff t$ and the precession rate $\sub{\dot\varphi}{lon}$. These measurements for the precession rate are superior to those from previous studies, since (i) Cepheid distances are much more accurate (at the relevant radii), (ii) no assumption is made regarding the stationarity of the warp (previous studies implicitly assumed $\diff i/\diff t=0$) and (iii) we do this over bins in guiding-centre radius $\sub{R}g$, thereby obtaining radial profiles for all these quantities.

This paper is organised as follows. In Section~\ref{sec:data}, we present the Cepheid sample, which we analyse in various ways in  Section~\ref{sec:analysis}. Finally, in Section~\ref{sec:discuss} we discuss our findings and conclude in Section~\ref{sec:conclude}.

\section{The data}
\label{sec:data}

\subsection{Cepheid characteristics}
We now briefly summarize some properties of Cepheids that have some relevance in the present context, but may not all be well known to our readers. Most Cepheids begin their life as B-type main-sequence stars. When such a star has become a red-giant (or even before it gets there) it ignites helium in the core and may perform a `blue loop' in the Hertzsprung-Russel diagram. On this loop it passes the instability strip on its way to higher temperatures and again on its way back to the asymptotic giant branch.

Whether a star undergoes such a blue loop and how long it takes, depends sensitively on its mass, metallicity, and helium abundance. At near-Solar metallicities, stars with mass 4.5-10\,\Msun will become Cepheids \citep{Anderson2014}, while at sub-Solar metallicities this range becomes 3-12\,\Msun, i.e.\ reaching into A-type stars. This metallicity dependence has two consequences. First, Cepheids are more abundant in metal-poor populations (owing to the scarcity of higher-mass stars) such as in the SMC, and in the Milky-Way disc their fraction increases outwards. Second, metal-poor Cepheids are on average fainter and older than metal-rich Cepheids. 

How long a star lives before its blue loop depends sensitively on its rotation \citep{AndersonEtAl2016A}, with fast rotators taking almost twice as long. Since binarity can sustain the initial fast rotation of B stars, and Cepheids are often found to be binary \citep{KervellaEtAl2019}, single-star evolutionary tracks may significantly underestimate their true age.

\subsection{The sample}
We use the sample of Milky-Way Classical Cepheids presented by (\citealt{Gaia22}, hereafter \citetalias{Gaia22}) that consists of Cepheids identified \citep{Ripepi2022_DR3} in Gaia DR3 \citep{Gaia:DR3} as well as those from the catalogues of \cite{Inno2021} and \cite{Pietrukowicz2021}. It combines EDR3 astrometry \citep[position, proper motion,][]{Lindegren21} and DR3 radial velocities  with distances obtained from the CC period-luminosity relation \citep{Ripepi2019,Ripepi22} to obtain the full 3D kinematics for 2123 stars\footnote{Adding near-infrared detected Cepheids can in theory extend the sample to behind the Galactic centre, providing a much wider view of the Milky-Way disc. However, as the danger of confusing spotted stars for Cepheids is substantial \citep{Pietrukowicz2015}, we refrain from such an extension to keep a homogeneous sample.}. We follow \cite{Semczuk22:gap} and adopt for the velocity of the Sun (needed to convert to Galacto-centric phase-space coordinates) 
\begin{align}
    \label{eq:Vsun}
    \vec{v}_\odot = (13,\,250,\,6.9)\,\kms,
\end{align}
which is a blend of the results derived by \cite{SBD2010, Schoenrich2012} and deviates from that adopted by \cite{Gaia22}. As these values are not very well known (systematics dominate over uncertainties) and affect our analysis, we will explore the effect of different values. Similarly, we explore the effect of an alternative Cepheid period-luminosity relation \citep{CRA2023}.


\begin{figure}
	\includegraphics[width=\columnwidth]{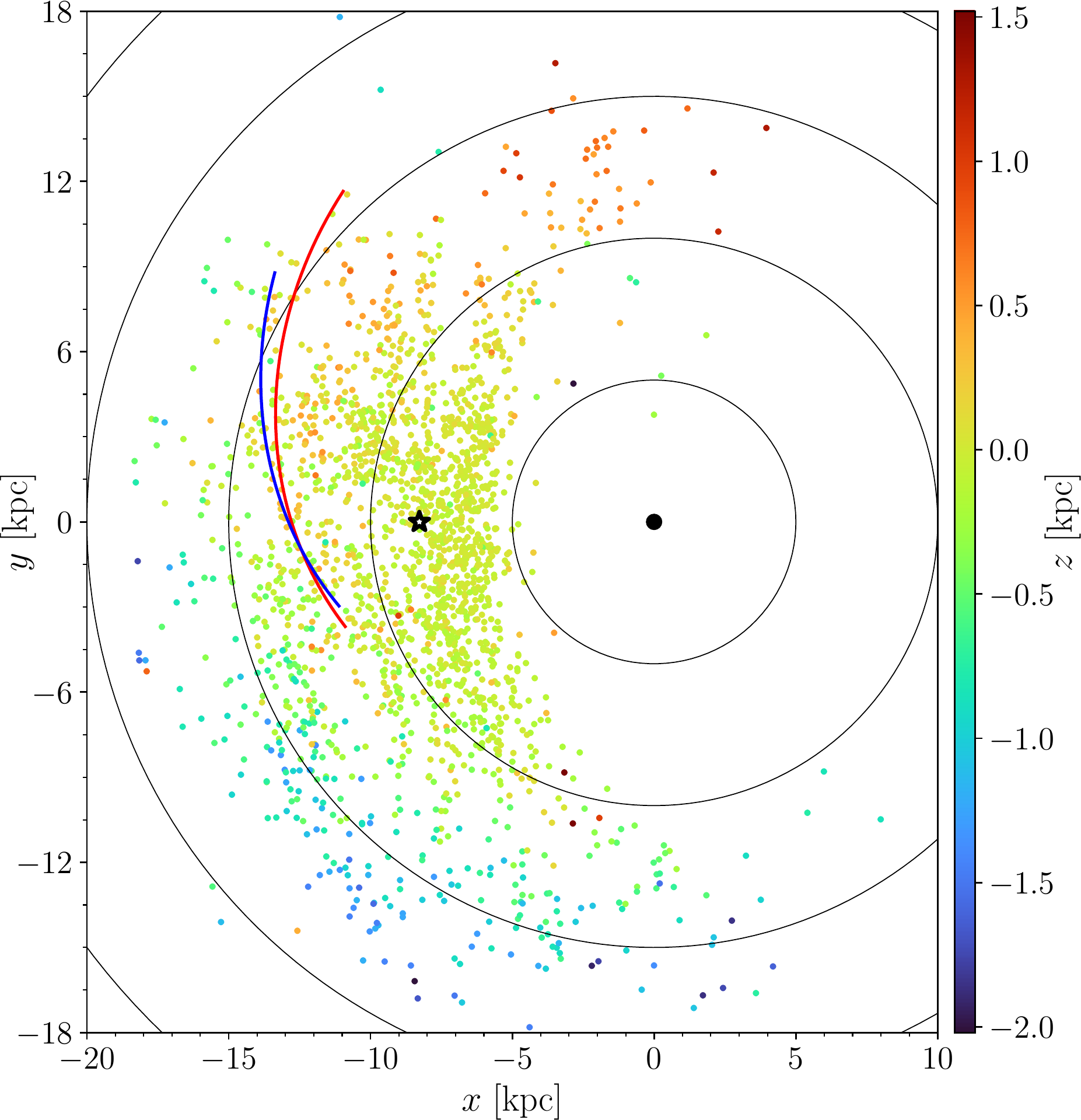}
    \vspace*{-5mm}
    \caption{
    The distribution in the Galactic plane (the Sun and Galactic centre are indicated, Galactocentric circles differ by 5\,kpc in radius) of the 2009 Classical Cepheids analysed (excluding 114 high-eccentricity stars, see text). Colour indicates vertical position (clipped to $[-2,1.5]$kpc). The arcs indicate the twisting line of nodes of the warp as found from our analyses of the Cepheid's orbital planes (red) and positions (blue).
    }
    \label{fig:XYZ}
\end{figure}

The sample contains some stars on significantly non-circular orbits, which could be artefacts of erroneous distance (due to mis-classification of a non-Cepheid) and/or radial velocity or could be caused by the kick suffered when a supernova explodes in a binary (of which the Cepheid progenitor was the secondary). Whatever the cause, eccentric orbits are also likely to be randomly inclined and hence no good tracers of any warp. We therefore remove stars with orbital eccentricity $e>0.2$, defined as the ratio between the epicycle amplitude and the radius of the guiding-centre orbit. We compute $e$ using the epicycle approximation as (see \citealt{BT2008}, equations 3.98 and 3.99)
\begin{align}
    \label{eq:e}
    e = \frac{\gamma}{2v_{\mathrm{circ}}} \left[\gamma^2(v_{\mathrm{t}}-v_{\mathrm{circ}})^2 + v_r^2\right]^{1/2}.
\end{align}
Here, $\gamma=2\Omega/\kappa$ with $\Omega$ and $\kappa$ the tangential and radial epicycle frequencies, $v_{\mathrm{t}}$ and $v_r$ are the star's tangential and radial velocity, and $v_{\mathrm{circ}}$ is the speed of the circular orbit with the same total angular momentum as the star. We assume a flat circular speed curve with $v_{\mathrm{circ}}=238$\,\kms, when $\gamma=\sqrt{2}$. This procedure removes 114 stars such that 2009 remain (our findings are  insensitive to the precise value of the eccentricity cut-off or the assumption of a flat circular-speed curve, as demonstrated in Section~\ref{sec:ana:alt}).

In Fig.~\ref{fig:XYZ}, we plot the distribution of these stars in the Galactic plane, indicating in colour their vertical offset from the mid-plane (as defined by Galactic latitude $b=0$).

As the Cepheids in the outer disc are generally metal poor compared to the Sun, they are also less massive (as explained above) and older. The ages estimated by \citetalias{Gaia22} for this sample range between 100\,Myr for Cepheids near the Sun to about 200\,Myr in the outer disc, although the true ages may well be up to twice as large owing to binarity, as explained above.

\begin{figure}
	\includegraphics[width=\columnwidth]{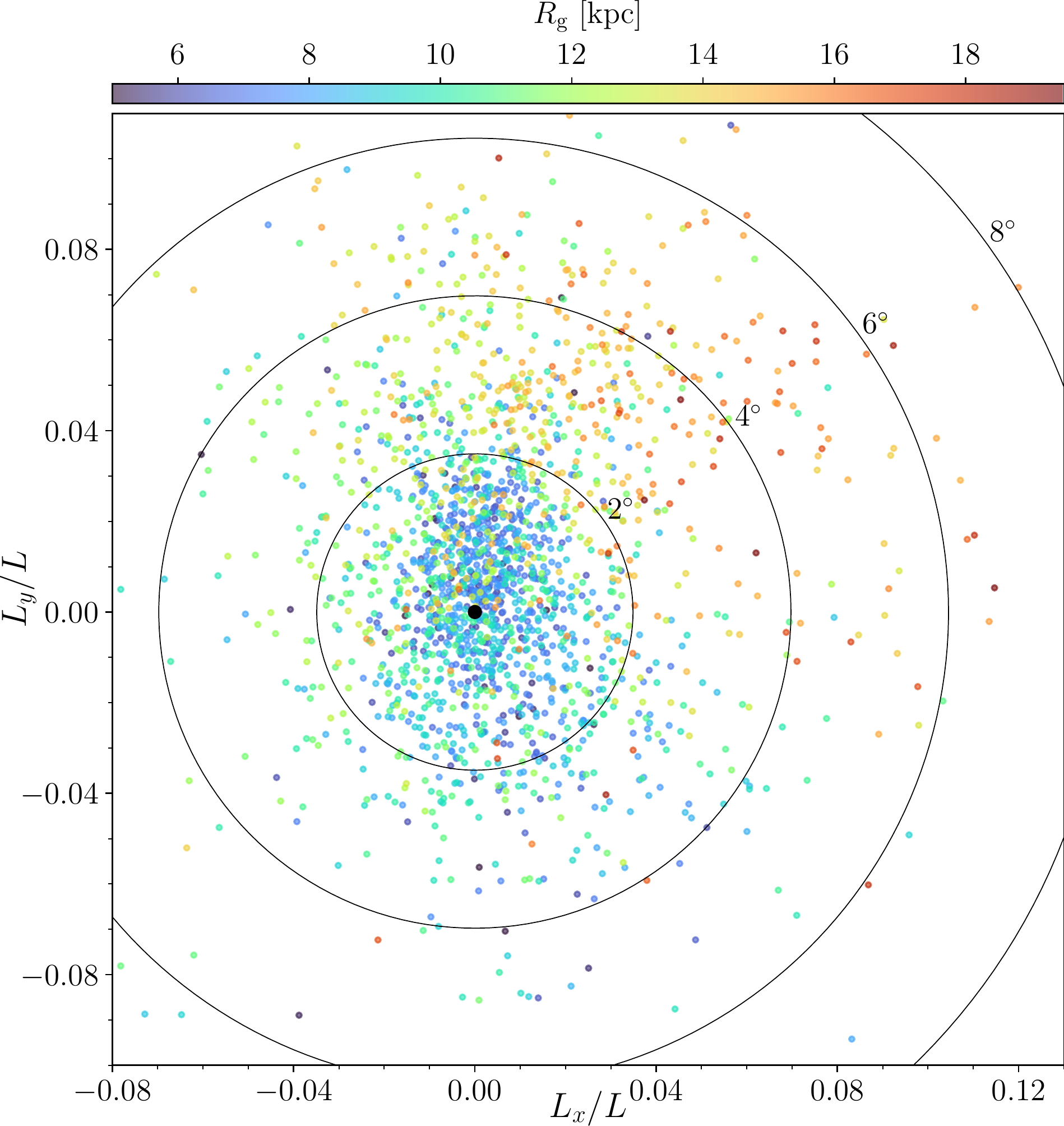}
    \vspace*{-5mm}
    \caption{
    The distribution of Galactic Cepheids from Fig.~\ref{fig:XYZ} over the $x$ and $y$ components of the directions $\hvec{L}$ of angular momentum. Colour indicates the guiding centre radius of their orbit. A value of $(\hat{L}_x,\hat{L}_y)=(0,0)$ (black dot) corresponds to alignment with Galactic coordinates, while circles indicate tilts of 2\degr, 4\degr, 6\degr and 8\degr.
    }
    \label{fig:LxLy}
\end{figure}

\section{Analysis}
\label{sec:analysis}

The Galactic warp is clearly visible in Fig.~\ref{fig:XYZ} as a systematic vertical displacement of the stars in Fig.~\ref{fig:XYZ}. Several recent studies of similar samples of Galactic Cepheids \citep{Chen2019, Skowron2019A, Skowron2019B, Lemasle2022} have analysed solely their spatial distribution and done so predominantly by fitting a parametrized radial warp profile. Here, we improve on these studies by (i) including the velocities in our analysis and (ii) performing it in radial bins (non-parametric). 

A better estimate for a star's typical radial position than its instantaneous radius is the guiding-centre radius $R_{\mathrm{g}}$ of its orbit, i.e. the radius of a circular orbit with the same angular momentum. We split the sample into 22 bins in $R_{\mathrm{g}}$ (computed using the same model for $\sub{v}{circ}$ as for the eccentricity cut) with the same number of stars per bin, and also consider the 21 intermittent bins between the medians of these primary bins, resulting in 43 bins in total.

\subsection{Instantaneous orbital planes}
Adding the velocities $\vec{v}$ to the positions $\vec{x}$ enables the computation of the angular momenta $\vec{L}=\vec{x}\times\vec{v}$ and hence of the instantaneous orbital planes perpendicular to $\vec{L}$. The unit vector $\hvec{L}=\vec{L}/|\vec{L}|$ defines for each star the unit \emph{normal} to or \emph{pole} of its instantaneous orbital plane. In Fig.~\ref{fig:LxLy}, we plot the distribution of the projection of these poles onto the $x$-$y$ plane 
(equivalent to pole-count maps or great-circle cell counts, \citealt{JohnstonEtAl1996}). Stars whose orbital plane is aligned with the Galactic coordinates would be at the origin in this plot, while $\hvec{L}$ on circles around the origin are inclined by the same degree, but in different direction.

\begin{figure*}
	\includegraphics[height=64mm]{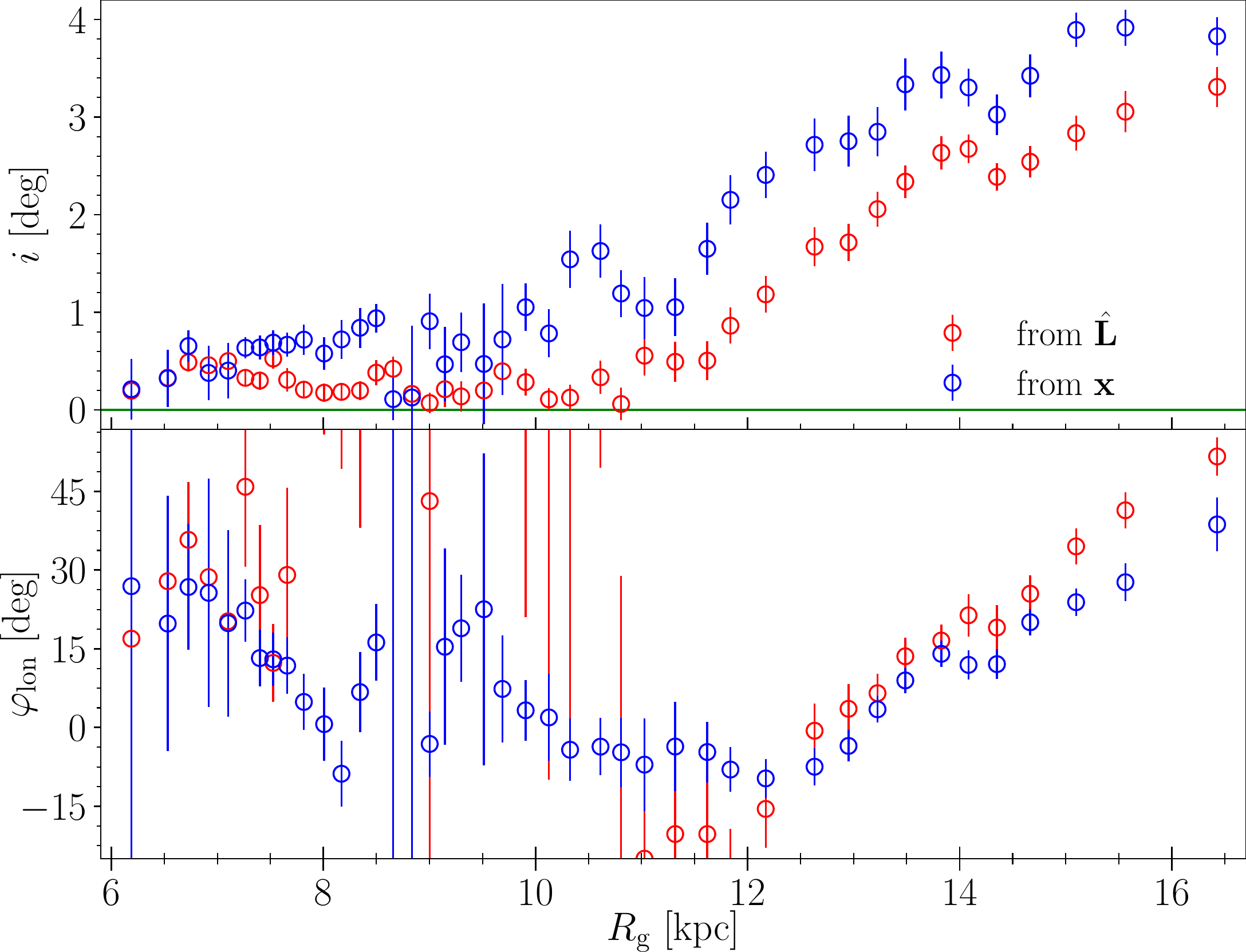}\hfill
    \includegraphics[height=64mm]{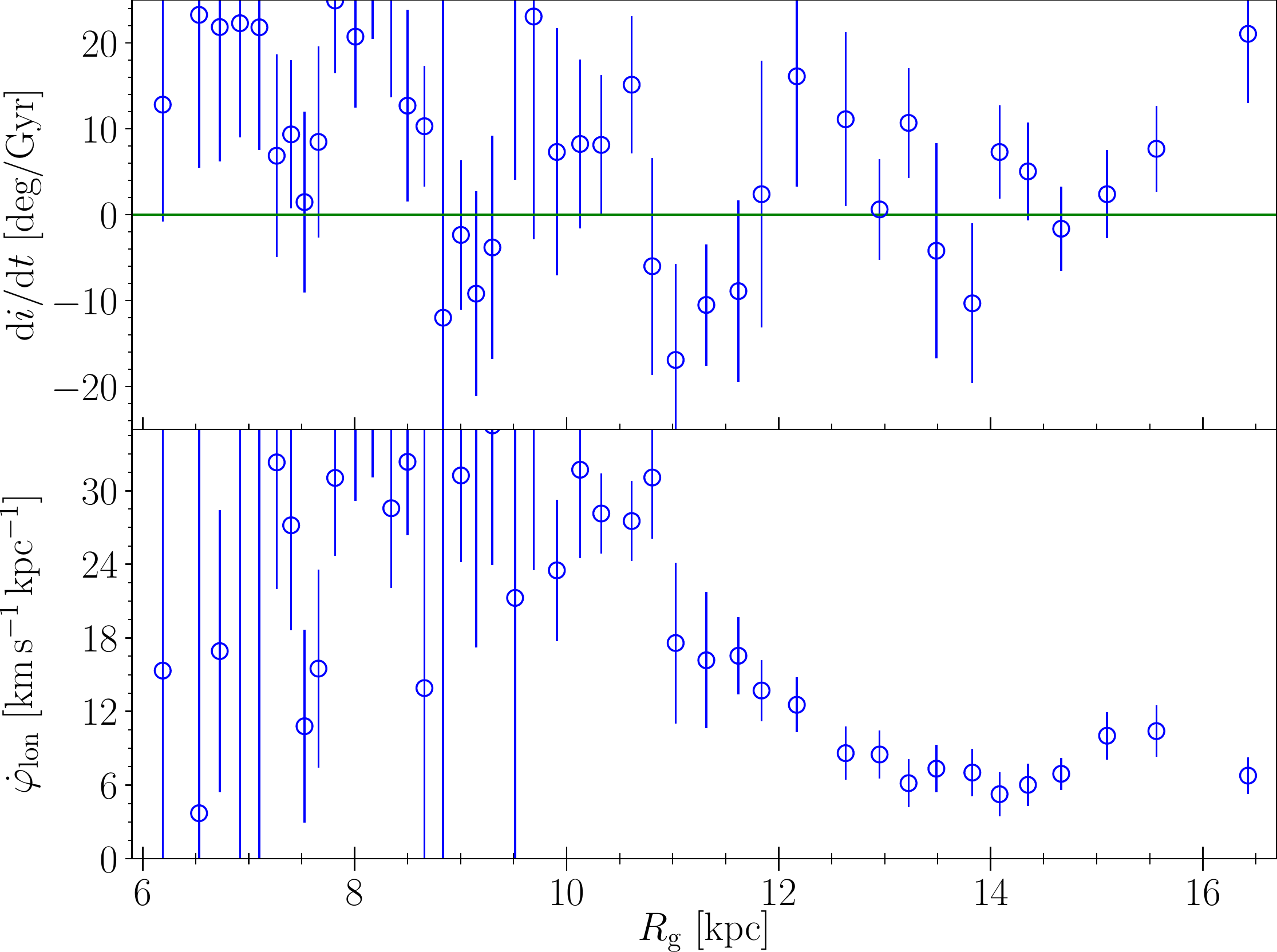}
    \caption{
    \textbf{Left}:
    tilt (inclination) $i$ (top) and azimuth $\varphi_{\mathrm{lon}}$ of the line of nodes (bottom, defined to be 0\degr\ for the position of the Sun and increasing in direction of Galactic rotation) for Cepheids in 43 bins in guiding-centre radius $R_{\mathrm{g}}$. The red points are obtained from the mean angular-momentum directions $\langle\hvec{L}\rangle$, while the blue points are obtained by fitting an inclined plane to the positions only. The large noise and uncertainties for $\varphi_{\mathrm{lon}}$ at small $\sub{R}g$ reflect the fact that $\varphi_{\mathrm{lon}}$ is ill-determined for small inclination.
    \textbf{Right}:
    The rate of change $\diff i/\diff t$ of the tilt (inclination, top) and the precession rate $\sub{\dot\varphi}{lon}$ for the inclined planes fitted to the Cepheids position in each of 43 bins in guiding-centre radius $\sub{R}g$.
    }
    \label{fig:inc,lon,dinc,omp}
\end{figure*}

In Fig.~\ref{fig:LxLy}, we colour code the stars according to $R_{\mathrm{g}}$. Stars at small $R_{\mathrm{g}}$ cluster more around $\hvec{L}=0$, reflecting a flat inner disc, while stars at larger $R_{\mathrm{g}}$ tend to have $\hat{L}_y>0$, indicative of the warp. In each of the 43 $\sub{R}g$ bins, we compute the mean $\langle\hvec{L}\rangle$\footnote{In practice, we compute the means of $\hat{L}_x$ and $\hat{L}_y$, whereby recursively ignoring stars outside of $3\sigma$ from the sample mean, though our findings are not sensitive to this $\kappa$-$\sigma$ clipping. The mean of $\hat{L}_z$ is then constructed from the normalisation condition.}, which corresponds to a mean inclination and a mean azimuth for the line of nodes, which we plot in the left panels of Fig.~\ref{fig:inc,lon,dinc,omp} (red circles) against the median guiding-centre radius for each bin. At $R_{\mathrm{g}}\lesssim11\,$kpc, the angular momenta of the Cepheids are on average close to being aligned with the Galactic south pole. The orbital planes of most stars in this inner part of the disc are inclined by $\lesssim1\degr$ but in different directions with some preference for $\sub{\varphi}{lon}\sim0\degr$ ($\hat{L}_y>0$) and to a lesser degree $\sim180\degr$. In other words, the inner disc at $R_{\mathrm{g}}\lesssim11\,$kpc does not show a coherent warp, but possibly a more complex picture (including a bent of the disc or a lop-sided warp). The situation is further complicated by possible systematic errors in the assumed Solar velocity and dust obscuration. We therefore refrain from a further analysis of the Cepheid warp at $R_{\mathrm{g}}\lesssim11\,$kpc.

At $R_{\mathrm{g}}\gtrsim11\,$kpc, on the other hand, the inclination increases quickly to 3\degr at 14\,kpc and appears to stay at that level towards larger radii, although our sample does not probe much beyond $R_{\mathrm{g}}=16\,$kpc. 

In this inclined outer part of the Cepheid disc, we can also measure the orientation of the warp as described by the azimuth $\sub{\varphi}{lon}$ of the ascending nodes, shown in the bottom left panel of Fig.~\ref{fig:inc,lon,dinc,omp}. We find that $\sub{\varphi}{lon}$ is in fact not constant, but increases with guiding-centre radius. This is in contrast to the usual modelling of the Cepheid warp in previous studies, all of which assume a single value (often set to $\sub{\varphi}{lon}=0\degr$). $\sub{\varphi}{lon}$ increasing with radius corresponds to a twisting warp with a leading spiral of nodes, which we indicate as red arch in Fig.~\ref{fig:XYZ}. We find that the dependence is very well described by a simple linear increase and hence naturally described by its gradient $\diff\sub{\varphi}{lon}/\diff\sub{R}g$. In the remainder of this study, we refer to this gradient as the \emph{twist} of the warp.

\subsection{Fitting inclined planes}
As our sample has no Cepheids behind the Galactic centre, and hence does not fully cover all azimuths, we cannot reliably distinguish the $m=1$ and $m=2$ Fourier components as defined in equation~\ref{eq:Z:Fourier}. Instead, we fit for each bin in $\sub{R}g$ a precessing inclined plane (which for small inclincation corresponds to the $m=1$ Fourier component) to the Cepheids' Galactocentric directions
$\hvec{x}\,=\vec{x}/|\vec{x}|$ and their time derivatives. To this end, we first find $\vec{a}\equiv\{a_x,a_y\}$ that minimises 
\begin{align} \textstyle 
    \chi^2=\sum_i (a_x\hat{x}_i + a_y\hat{y}_i-\hat{z}_i)^2,
\end{align}
which is $\vec{a}=\tens{A}^{-1}\cdot\vec{b}$ with matrix $\tens{A}=\sum_i\{\hat{x}_i,\hat{y}_i\}\otimes\{\hat{x}_i,\hat{y}_i\}$ ($\otimes$ denotes the outer product) and vector $\vec{b}=\sum_i\hat{z}_i\{\hat{x}_i,\hat{y}_i\}$. Second, we obtain the normal to the inclined plane as $\hvec{n}=\{a_x,a_y,-1\}(a_x^2+a_y^2+1)^{-1/2}$. Third, we compute $\dot{\tens{A}}$ and $\dot{\vec{b}}$ from the velocities and obtain $\dot{\vec{a}}=\tens{A}^{-1}\cdot(\dot{\vec{b}}-\dot{\tens{A}}\cdot\vec{a})$. Finally, the inclination and azimuth of the line of nodes are obtained as $i=\cos^{-1}|\hat{n}_z|$ and $\sub{\varphi}{lon}=\tan^{-1}(\hat{n}_x/\hat{n}_y)=\tan^{-1}(a_x/a_y)$ and their time derivatives by straightforward differentiation.

\begin{figure*}
	\includegraphics[width=145mm]{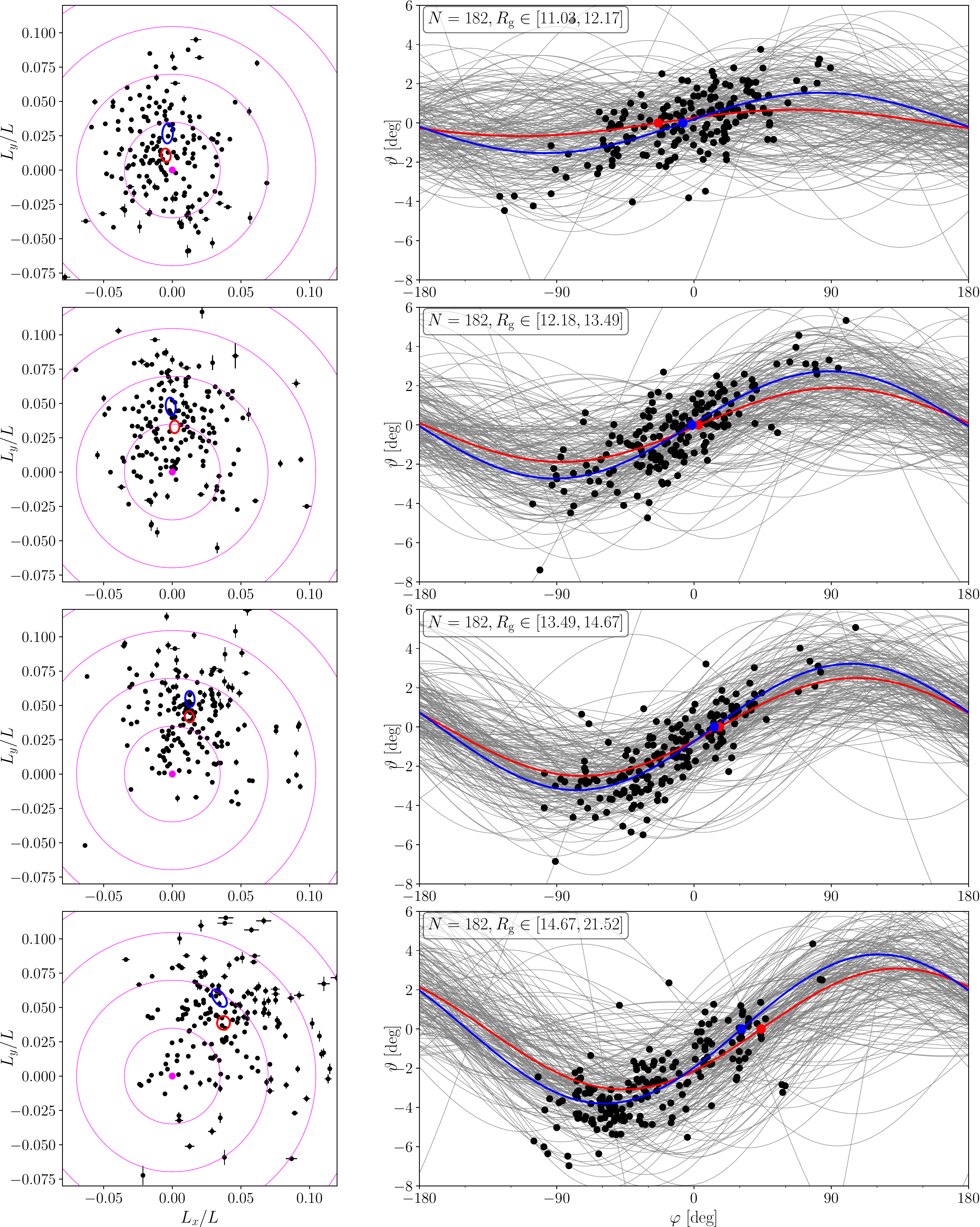}
    \caption{
    Warp for Cepheids in four distinct bins with increasing (from top to bottom) guiding-centre radius $R_{\mathrm{g}}$, as indicated in the legends (each of these bins combines two of the 22 bins used for analysis). The \textbf{left} panels are like Fig.~\ref{fig:LxLy}, but include individual error bars. The \textbf{right} panels show the spatial distribution of Cepheids (dots) over Galactocentric azimuth $\varphi$ and latitude $\vartheta$ (in these coordinates the Sun is at the origin and rotation to the right) and for each the projection of its instantaneous orbital plane (grey). The mean $\hvec{L}$ is indicated in the left panels as 2-$\sigma$ (86 per cent) confidence region (red) and in the right panels as projection of the normal plane (red). Similarly, the blue error ellipse (left panel) and curve (right panel) reflect the tilted plane that best fits the Cepheid positions.
    }
    \label{fig:planes}
\end{figure*}

\begin{table*}
    \caption{Dependence of the results for the twist $\diff\sub{\varphi}{lon}/\diff\sub{R}g$ and the precession rate $\sub{\dot\varphi}{lon}$ at $\sub{R}g=12$ and 14\,kpc (from bins with 100 Cepheids on either side of these) on eccentricity cut, the $\kappa$-$\sigma$ clipping, the assumed Solar velocity, the number of bins, the assumed rotation curve, and the Cepheid $P$-$L$ relation. Values/methods not mentioned in the first column are at default (see Section~\ref{sec:data}).}
    \label{tab:results}
    \centering
    \begin{tabular}{lllll}
        deviation & 
        \multicolumn{2}{c}{$\diff\sub{\varphi}{lon}/\diff\sub{R}g$ [deg/kpc]} &
        \multicolumn{2}{c}{$\sub{\dot\varphi}{lon}$ [\kmskpc]}
        \\
        from default & from $\hvec{L}$ & from $\hvec{x}$
        & at $\sub{R}g=12\,$kpc & at $\sub{R}g=14\,$kpc 
        \\
        \hline
        default ($\vec{v}_{\odot}=(13,250,6.9)\,$\kms, $e_{\max}=0.2$) & $14.7\pm0.7$ & $10.6\pm0.8$ & $12.4\pm1.6$ & $5.9\pm1.3$\\
        maximum eccentricity $e_{\max}=0.1$ & $15.4\pm1.0$ & $11.6\pm1.2$ & $12.7\pm1.5$ & $6.1\pm1.0$
        \\
        maximum eccentricity $e_{\max}=0.3$ & $14.6\pm0.7$ & $10.5\pm0.9$ & $13.1\pm1.7$ & $6.2\pm1.3$
        \\
        using 44/87 primary/total bins & $14.1\pm0.7$ & $10.4\pm0.7$ \\
        no $\kappa$-$\sigma$ clipping to find $\langle\hvec{L}\rangle$ & $14.9\pm0.7$ & 
        \\
        $\vec{v}_{\odot}=(9.3,251,8.6)\,$\kms\,(\citetalias{Gaia22}) & 
            $13.0\pm0.7$ & $10.7\pm0.8$ & $\phantom{0}8.9\pm1.7$ & $3.9\pm1.4$\\
        $v_{\odot,z}=8.6\,$\kms & $13.6\pm0.6$ & $10.7\pm0.8$ & $\phantom{0}8.9\pm1.6$ & $3.7\pm1.4$\\
        $v_{\odot,R}=9.3\,$\kms & $14.3\pm0.8$ & $10.9\pm0.8$ & $12.0\pm1.6$ & $6.1\pm1.3$\\
        $v_{\odot,\varphi}=246\,$\kms & $14.3\pm0.7$ & $11.2\pm0.9$ & $10.8\pm1.5$ & $6.0\pm1.2$\\
        $v_{\odot,\varphi}=254\,$\kms & $15.2\pm0.9$ & $10.4\pm1.0$ & $14.7\pm1.7$ & $7.0\pm1.3$\\
        $\sub{v}{circ}=238\,\kms\,(R/R_0)^{-0.05}$ & $14.0\pm0.6$ & $10.1\pm0.8$ & $14.3\pm1.8$ & $6.5\pm1.2$\\
        Cepheid $P$-$L$ relation from \cite{CRA2023} &
        $14.9\pm0.8$ & $10.6\pm0.9$ & $12.4\pm1.5$ & $5.9\pm1.3$
    \end{tabular}
\end{table*}

\subsubsection{Inclination and line of nodes}
In Fig.~\ref{fig:planes}, we plot for four distinct bins in $\sub{R}g$ (increasing from top to bottom) the distributions over $\hat{L}_x,\,\hat{L}_y$ (as in Fig.~\ref{fig:LxLy}) in the left panels, while the right panels show the corresponding individual orbital planes (grey curves) projected onto Galactocentric azimuth and latitude (such that the Sun is at the origin and rotation to the right). These orbital planes show a clear and coherent warp, which is in good agreement with the spatial distribution (dots), at least at $\sub{R}g\gtrsim12\,$kpc. We show in red the average orbital plane (in the sense of the average $\langle\hvec{L}\rangle$) and in blue the inclined plane fitted to the normalised positions. In left panels of Fig.~\ref{fig:inc,lon,dinc,omp} we also plot (blue) the tilt (inclination $i$) and direction $\sub{\varphi}{lon}$ of the line of nodes for the fitted inclined planes.

The inclined planes fitted to the positions are generally slightly more inclined than the average of the individual orbital planes. This could be (partly) caused by observational biases (since obscuration by dust hides some stars at low $|z|$) or be a real effect.

\subsubsection{Tumble and precession}
In the right panels of Fig.~\ref{fig:inc,lon,dinc,omp}, we plot the rates $\diff i/\diff t$ and $\sub{\dot{\varphi}}{lon}$ against $\sub{R}g$. At $\sub{R}g\lesssim11\,$kpc, the warp amplitude is too small for reliably measuring these rates (resulting in large scatter and uncertainties). The rates of change of the inclination are consistent with zero for most of the bins, except perhaps the marginally significant positive values for the outermost bin.

The precession rates, however, clearly show fast prograde precession, with $\sub{\dot\varphi}{lon}\approx12\,\kmskpc$ at $\sub{R}g=12\,$kpc, decreasing to $\sim6\,\kmskpc$ at $\sub{R}g\gtrsim13\,$kpc. This is in clear contradiction to the slow retrograde precession expected for a simple warp, but the former of these values is in good agreement with previous estimates for the general stellar population \citep{Poggio20,Cheng20}, which mainly probes $R\sim12\,$kpc. Our measurements at $\sub{R}g\gtrsim 13\,$kpc are significantly smaller, but still positive and inconsistent with the expectation of a simple warp.

\subsection{Effects of the Solar velocity and other parameters}
\label{sec:ana:alt}
We repeated the analyses leading to Fig.~\ref{fig:inc,lon,dinc,omp} for twice the number of (primary) bins in $\sub{R}g$, no $\kappa$-$\sigma$ clipping for finding $\langle\hvec{L}\rangle$, as well as various deviating choices of the Solar velocity, the threshold for the eccentricity cut, the assumed circular speed curve (affecting eccentricity $e$ and guiding centre $\sub{R}g$), and the Cepheid period-luminosity relation. For the latter we explored the recent relation by \cite{CRA2023}, which is based on the metallicity-effect calibration by \cite{BreuvalEtAl2022}. For $\vec{v}_\odot$, we explored the values adopted by \cite{Gaia22}, either for all components or for the radial and vertical component only (which have the strongest effect).

The resulting profiles (not shown) for inclination $i$, line-of-node orientation $\sub{\varphi}{lon}$, and their time derivatives are largely similar to those obtained from our default values shown in Fig.~\ref{fig:inc,lon,dinc,omp}, including the onset of the warp at $\sub{R}g\approx11\,$kpc, the typical inclination of $i=3$-4 degrees in the outer disc, and the fact that $i$ found from orbital-plane analysis is generally $\sim1\degr$ smaller than that found from fitting a single plane to all Cepheids in a given $\sub{R}g$ bin. The largest variation is found in the orientations $\sub{\varphi}{lon}$ of the outer disc and its rate of change $\sub{\dot\varphi}{lon}$, as summarized in Table~\ref{tab:results}. Generally the twist $\diff\sub{\varphi}{lon}/\diff\sub{R}g$ is 10-11  deg/kpc for the single-plane fits and 13-15 deg/kpc for the average orbital plane, with statistical uncertainties $\lesssim1$deg/kpc.

Obviously, different assumptions for the Solar velocities have the strongest effect on $\diff\sub{\varphi}{lon}/\diff\sub{R}g$ obtained from $\langle\hvec{L}\rangle$ and on the precession rates $\sub{\dot\varphi}{lon}$, both of which depend on the Cepheid velocities. Assuming a larger radial and/or smaller vertical velocity for the Sun (as did \citetalias{Gaia22}), reduces both of these measures with the strongest effect on $\sub{\dot\varphi}{lon}$ at $\sub{R}g=12\,$kpc. However, while the changes are significant (larger than the statistical uncertainties), they are not substantial: the twist inferred from the individual orbital planes is still larger then that obtained from a inclined-plane fit and the precession remains fast and prograde. These results are hence robust.

\begin{figure}
	\includegraphics[width=\columnwidth]{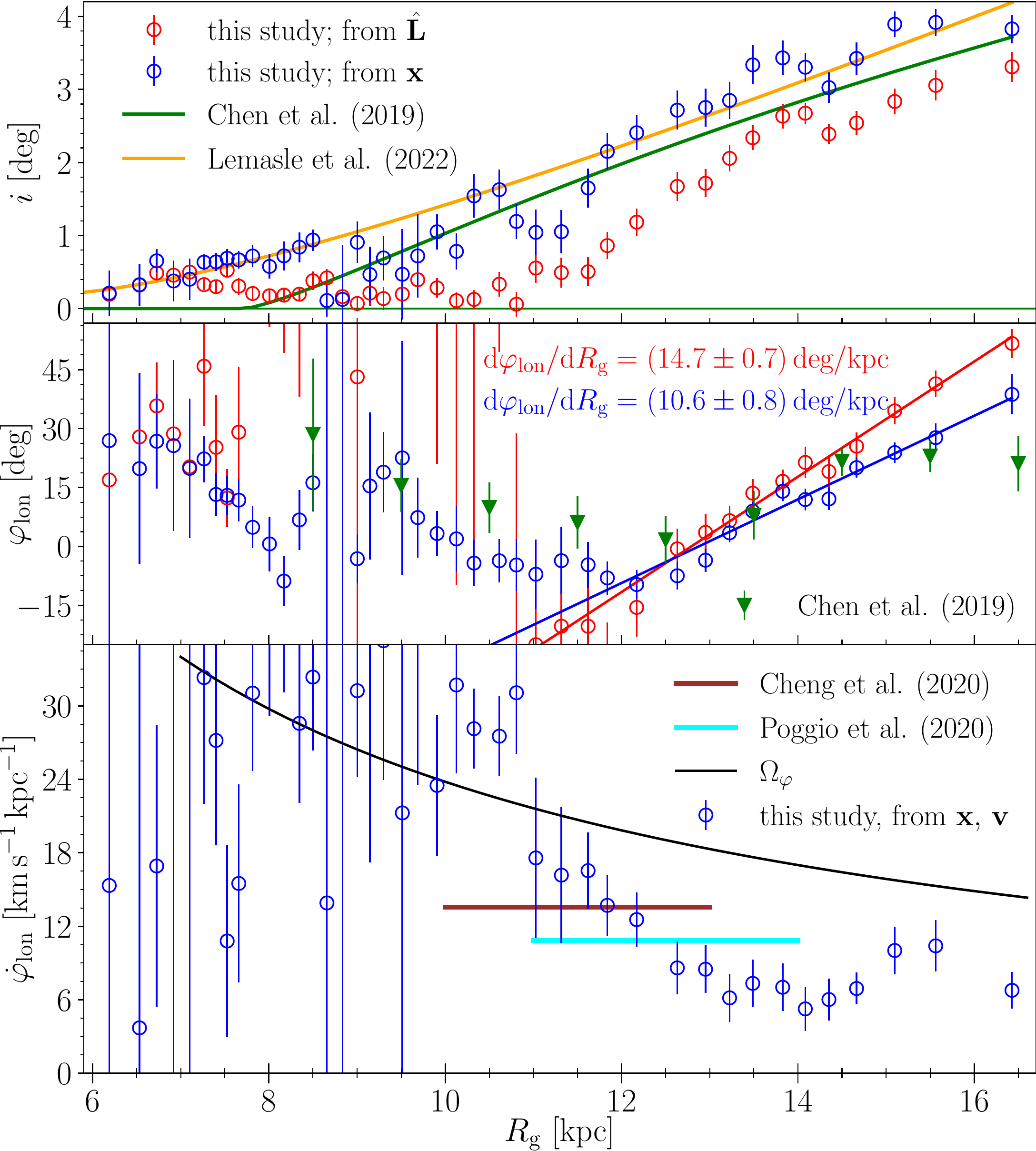}
    \vspace*{-5mm}
    \caption{
    The results from this study obtained by analysing the Cepheids' orbital planes (red circles) or fitting a precessing inclined plane (blue circles) to their positions and velocities are compared to previous studies of the Cepheid warp \citep{Chen2019,Lemasle2022} and the warp precession rate inferred from the vertical proper motion of giants \citep{Poggio20} or the general stellar population \citep{Cheng20}. As these latter two studies considered a single value for the precession rate at all radii (and also $\sub{\varphi}{lon}=0\degr$), we show their result over a radial range where the respective data sets carry most warp signal. We also plot $\Omega_\varphi$ for an assumed flat rotation curve.
    }
    \label{fig:inc,lon,omp}
\end{figure}

\begin{figure}
	\includegraphics[width=\columnwidth]{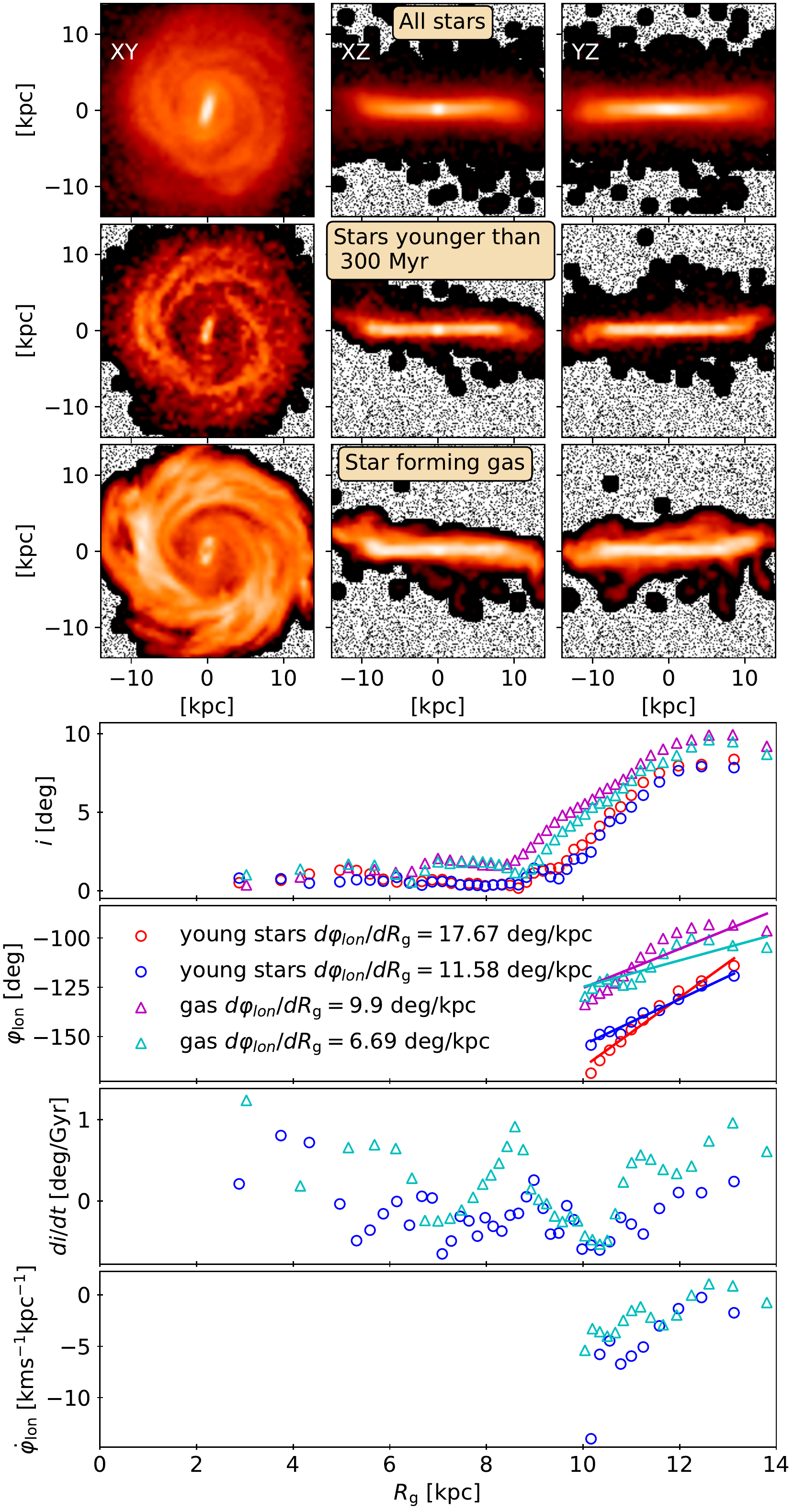}
    \vspace*{-5mm}
    \caption{
    \label{fig:TNG}
    Projected density (top) and warp parameters (bottom) determined for young star particles and star-forming gas of a simulated spiral (galaxy ID 566365 at $z=0$ of run TNG50-1) in the same ways as for the Milky-Way Cepheids (red/purple symbol refer to the analysis of orbital angular momenta, while blue/cyan refers to inclined planes fitted to the positions).
    }
\end{figure}

\section{Discussion}
\label{sec:discuss}
\subsection{Comparison to other studies}
Fig.~\ref{fig:inc,lon,omp} compares our results (as already presented in earlier figures) to previous studies of the inclination and line of nodes of the Cepheid warp, as well as for precession rate of the general stellar warp. Our results largely agree with those previous studies, but our analysis provides a more detailed view, largely owed to our binning of the Cepheid in guiding-centre radius $\sub{R}g$ as opposed to an overall fit or a global model. The only previous exception to this global-fit approach was an analysis using radial bins by \cite{Chen2019}, which allowed them to measure the radial change of the line of nodes (green triangles in the middle panel of Fig.~\ref{fig:inc,lon,omp}). Their result agrees mostly with ours in the sense that the line of nodes forms a leading spiral at $R\gtrsim12\,$kpc, but we find the linear increase of $\sub{\varphi}{lon}$ with radius to continue all the way to our outermost data point at 16\,kpc, while \cite{Chen2019} find it to remain constant for $R\gtrsim14\,$kpc. Since our Cepheid sample extends further than that used by \cite{Chen2019}, our result appears more reliable.

Previous estimates \citep{Poggio20,Cheng20,Wang2020,Chrobkova2021} for the warp precession rate have been obtained from general stellar samples using distance information that was (at least partly) based on astrometric parallaxes with rather low accuracy and by \emph{assuming} that the warp is a coherently precessing (but otherwise unevolving) inclined disc with the same orientation ($\sub{\varphi}{lon}$) and precession ($\sub{\dot\varphi}{lon}$) at all radii. None of these assumptions is made by our analysis, which allows the inclination of the warp to vary with time and its orientation and precession rate with radius. Moreover, while the Cepheids sample is much smaller than the large samples used in those previous studies, their distances are excellent, preventing biases from distance systematics. Our measurements of $\sub{\dot\varphi}{lon}$ (bottom panel of Fig.~\ref{fig:inc,lon,omp}) show a fast and radially decreasing precession, consistent with the previous estimates by \cite{Poggio20, Cheng20}, but not with those of \cite{Wang2020, Chrobkova2021}, who found $\sub{\dot\varphi}{lon}\sim0$. It is not entirely clear, what drives these differences (also between those previous studies)\footnote{\cite{Chrobkova2021} claim that unsuitable warp parameters used by \cite{Poggio20} invalidate their result, but \cite{Cheng20} fitted those parameters to the kinematics and found essentially the same result.}, but as our method is based on accurate distances and involves no model assumptions, it is more reliable than any of these previous estimates.

After submission of our work, we became aware of a preliminary analysis by Cabrera-Gadea et al.\ (in preparation) of the Cepheid warp (albeit using incomplete radial velocity information) with similar findings to ours regarding the twist and precession.

\subsection{Where does the Cepheid warp start?}
For each bin in guiding-centre radius $\sub{R}g$, we find a tilted plane describing the Cepheid warp in two different ways: as that normal to the mean angular momentum direction and by fitting to the positions, see Fig.~\ref{fig:inc,lon,omp}. The resulting tilt angles, in particular those obtained by the fits, agree well with those found in previous studies of the Cepheid warp \citep{Chen2019, Skowron2019A, Skowron2019B, Lemasle2022}, as expected. The tilt angles obtained from the average orbital planes are generally smaller by about one degree. This could be at least partly caused by dust hiding Cepheids in the Galactic mid-plane, such that our sample is biased towards those warped out of the plane. We suspect that this mechanism is responsible for the measured inclination of $\sim1\degr$ at $\sub{R}g\lesssim11\,$kpc, where the orbital plane analysis suggests a flat disc instead. This is akin to the situation with Galactic OB stars, whose velocities are well described by dynamic equilibrium models but not their positions, which is best explained by unknown dust obscuration \citep{LiBinney2022}.

If this explanation is correct, it implies that the Cepheid warp actually begins at about 11\,kpc and that any previously detected Cepheid warp inside that radius is an artefact of selection effects caused by dust obscuration\footnote{This is suggestion is corroborated by our (preliminary) analysing simulated warped galaxies in the TNG50 simulations, as we did not find one for which the orbital plane analysis obtains $i\sim0\degr$ while the plane-fitting gives $i\gtrsim1\degr$, as we found for the Milky Way inside $\sim11\,$kpc.}. This would also imply that any measurements for the line of nodes in this region are void, such as the trailing spiral, also reported by \citeauthor{Chen2019} (\citeyear{Chen2019}, shown by green triangles in the middle panel of Fig.~\ref{fig:inc,lon,omp} and consistent with our measurements).

\subsection{The warp line of nodes}
Another clear difference between the warp parameters obtained from the two methods is the twist of the line of nodes, which for the orbital-plane analysis is stronger (larger $\diff\sub{\varphi}{lon}/\diff\sub{R}g$). Apart from dust obscuration discussed above, there are other possible reasons for such differences. Even if the Cepheids are distributed in tilted rings, their angular momenta may well deviate from the normal to that ring, already because of the nodal precession. If the Cepheid warp is lopsided (like the \ion{H}{I} warp), more deviations are expected, which may also depend on azimuth, of which our sample covers only half of the full range, as Cepheids on the other side of the Milky Way are hidden by dust.

Such deviations between these two different analysis methods are actually common for simulated galaxies. We applied our methods to Cepheid tracers (star particles younger than 300\,Myr) in several simulated spiral galaxies of the IllustrisTNG (TNG50, \citealt{Pillepich2019, Nelson2019}) simulation and found a variety of behaviours, including a case that is similar to the Milky Way (shown in Fig.~\ref{fig:TNG}) with regard to the behaviour of $\diff\sub{\varphi}{lon}/\diff\sub{R}g$. 

\subsection{The warp precession rate}
The line of nodes of the Cepheid warp, being a leading spiral, adheres to Briggs' rule, i.e.\ behaves as most warped galaxies do and indeed also all those simulated galactic warps in TNG50 which we analysed. The standard theoretical explanation, as already mentioned in the introduction, is that the warp precesses slowly in a retrograde sense with the orbital precession rate, i.e.\ $\sub{\dot\varphi}{lon}=\Omega_\varphi-\Omega_z$, when faster precession at smaller radii results in a leading spiral. And indeed, of the simulated warped galaxies in TNG50 for which we checked this, all adhere to this explanation, at least in the sense that our measurement for $\sub{\dot\varphi}{lon}$ mostly obtains negative values at 0-10\,\kms, including the example presented in Fig.~\ref{fig:TNG}. However, our measurements for the Cepheids do not follow this trend, but instead obtain $\sub{\dot\varphi}{lon}>0$. Moreover, the radial decline (from 12 to 6\,\kmskpc between 12 and 14\,kpc) would wind the line of nodes into a \emph{trailing} spiral or, equivalently, unwind the observed leading spiral in only $\sim100\,$Myr.

This contradiction implies immediately, that the instantaneous $\sub{\dot\varphi}{lon}$ that we (and others) have measured cannot be the predominant warp precession rate for most of the history of the warp. 

A warp may in fact be a superposition of several bending waves with azimuthal wave number $m$, as stipulated in equation~\eqref{eq:Z:Fourier}. In this case, each mode precesses with rate
\begin{align}
    \label{eq:omega:m}
    \dot{\psi}_m \equiv \omega_m = \Omega_\varphi \pm \frac1m \Omega_z,
\end{align}
where the `+' sign obtains a fast and the `$-$' sign a slow wave (see \citealt{BT2008}, chapter 6.6). The fast waves tend to wind up quickly such that eventually only the slowest survives, which is the slow $m=1$ wave (`simple' warp), which precesses with the same rate as stellar orbits. However, lopsidedness (as observed for the Milky-Way $\ion{H}{I}$ warp, but not easily detectable for Cepheids and other stellar samples due to the limited azimuthal coverage), requires an $m=2$ mode. In fact, the slow $m=2$ mode precesses with $\Omega_\varphi - \tfrac12 \Omega_z \sim \tfrac12 \Omega_\varphi$, close to our measurement for $\sub{\dot\varphi}{lon}$. 

Previous studies estimated the precession rate assuming that the warp is an $m=1$ wave. Our method of fitting a inclined plane, while allowing for a change of warp amplitude, also ignores lopsidedness ($m=2$). However, as all of the available samples have rather limited azimuthal coverage, these modes cannot (currently) be easily disentangled. Instead, all these various analyses essentially measure the local propagation rate $\sub{\dot\varphi}{lon}$ of the combined warp (not of a single mode), since the line of nodes is close to $\varphi=0$ and well covered by these samples. In the presence of both $m=1$ and $m=2$ modes, i.e.\ for a lopsided warp, such measurement can in principle obtain anything between the precession rates of these modes, i.e.\ $\Omega_\phi-\Omega_z \le \sub{\dot\varphi}{lon} \le \Omega_\phi+\tfrac12\Omega_z$, depending on the phases and relative amplitude of the modes.

A warp can be excited either abruptly by galactic interaction(s), or gently by the continued accretion of misaligned material, which either directly forms the warp or generates an external torque that subsequently causes the warp. In the latter case, the warp tends to be dominated by the `simple' $m=1$ mode \citep[e.g.][]{OstrikerBinney1989, ShenSellwood2006} at all times. In the former case of a more abrupt excitation, all modes are initially excited. Over time, all but the `simple' $m=1$ mode decay, resulting in a flaring of the warped disc (\citealt{Poggio2021}; Moetazedian et al., in preparation). In the transient phase during this decay the actual line of nodes propagates in an erratic way (in fact there may be more than two nodes) where periods with $\sub{\dot{\varphi}}{lon}>0$ and $\sub{\dot{\varphi}}{lon}<0$ alternate, even if in the long term the average precession is retrograde (e.g.\ Fig.~7 of Moetazedian et al., in preparation).

The situation is more complex for gaseous warps, for which the hydrodynamical forces may result in a faster decay to the simple warp. Since the Cepheids have only been born from the gas about one half orbital period ago their warp should largely reflect that of the gas.

\section{Conclusion}
\label{sec:conclude}
We have analysed the sample of all Milky-Way classical optical Cepheids with Gaia DR3 radial velocity in terms of the Galactic warp. Unlike most previous studies, our analysis avoids fitting parameterised warp profiles, but instead we separate stars into bins according to their guiding-centre radius $\sub{R}g$, which is generally a better indicator of the star's typical Galactocentric radius than its actual instantaneous radius. Our main results are as follows.

\begin{itemize}
    \item At $\sub{R}g\lesssim11\,$kpc, the individual orbital planes, defined as normal to the Cepheid's angular-momentum vectors, are on average flat, i.e.\ not warped. This is in contrast to some warping by $1\degr$ obtained by fitting an inclined plane to the positions, which we suggest is caused by dust obscuration.
    \item At $\sub{R}g\gtrsim11\,$kpc, both the angular-momentum analysis and the fitted inclined planes show a warp with amplitude $3\degr$-$4\degr$. The line of nodes twists in the direction of rotation from $\sub{\varphi}{lon}=-15\degr$ at 12\,kpc to ~45\degr at 16.5\,kpc. This is well described by a linear increase and corresponds to a leading spiral, adhering to Briggs' rule for spiral galaxies. This result is not entirely novel, but our detection of the twist is much clearer than from previous studies \citep{Chen2019}.
    \item We also determine the rate of change of the inclined planes fitted to the Cepheid positions. We find no significant time derivative for the inclination (warp amplitude), but a clear signal for $\sub{\dot{\varphi}}{lon}$, the nodal precession rate of the inclined planes. We find $\sub{\dot{\varphi}}{lon}=(12.4\pm1.6)\,\kmskpc$ at $\sub{R}g=12\,$kpc, which broadly agrees with previous estimates from older stars. We also find $\sub{\dot{\varphi}}{lon}$ to decrease radially to $(5.9\pm1.3)\,\kmskpc$ at $\sub{R}g=14\,$kpc and remaining roughly at this level beyond that. These values are unexpected for an ordinary warp corresponding to a inclined disc in near-circular motion, for which $\sub{\dot{\varphi}}{lon}<0$. When acting over $\sim100\,$Myr this measured differential precession would unwind the leading spiral of nodes, strongly suggesting that the instantaneous prograde precession is only transient and that the Milky-Way warp is more complex.
    
    For a warp excited by satellite interactions, Moetazedian et al. (in preparation) find significant fluctuations in the warp $\sub{\varphi}{lon}$ (and hence by implication $\sub{\dot\varphi}{lon}$), which spike in amplitude after each interactions and subsequently decline. This suggests that the measured $\sub{\dot\varphi}{lon}>0$ relates to a recent interaction, most likely with the Sgr dwarf galaxy, the only recent close interactor of the Milky Way.
    
\end{itemize}

\section*{Acknowledgements}
We are grateful to R.~Drimmel and S.~Khanna for providing the ages and distances of classical Cepheids from \citetalias{Gaia22}. We thank Richard Anderson for providing his improved Cepheid period-luminosity relation prior to publication. We appreciate insightful discussions with Lia Athanassoula, Andreas Just, Peter Berczik, and Hossam~Aly, and thank the reviewer, R.~Drimmel, for a prompt and useful report. This work was supported by STFC grant ST/S000453/1. RS thanks the Royal Society for generous support via a University Research Fellowship. This work has made use of data from the European Space Agency (ESA) mission Gaia (\url{https://www.cosmos.esa.int/gaia}), processed by the Gaia Data Processing and Analysis Consortium (DPAC, \url{https://www.cosmos.esa.int/web/gaia/dpac/consortium}). Funding for the DPAC has been provided by national institutions, in particular the institutions participating in the Gaia Multilateral Agreement.

\section*{Data Availability}
No data were generated in this study.


\bibliographystyle{mnras}
\bibliography{main} 

\bsp	
\label{lastpage}
\end{document}